# Study on Relative Collection Efficiency of PMTs with Point Light

Hai-qiong Zhang[a,b]   Zhi-min Wang[b)]   Feng-jiao Luo[b,c]   An-bo Yang[a,b,c]   Zhong-Hua Qin[b,c]
Chang-gen Yang[b]   Yue-kun Heng[a,b,c]

a. University of Chinese Academy of Science, Beijing 100049, China
b. Institute of High Energy Physics, Beijing 100049, China
c. State Key Laboratory of Particle Detection and Electronics, Beijing 100049, China

**Abstract:** Photon detection efficiency is a key parameter of PMTs in high-precision neutrino and dark matter experiments, while most of these experiments are focus on quantum efficiency. More and more experiments are trying to know the detection efficiency for the simulation of the detector such as JUNO. In order to have a good understanding on photon detection efficiency of the large-area PMT , we conducted a detailed comparison of the relative collection efficiencies of a series of dynode PMT with different sizes and collector structures. This study is based on the tests of relative quantum efficiency and relative detection efficiency with cross check by several light intensities. The testing and results will be discussed.

## 1.Introduction

Photomultiplier tube (PMT) is an important sensor to count photons and to measure parameters in particle physics experiments. Photon detection efficiency (DE) of PMT is always a key parameter for detector response while DE is a comprehensive effect of quantum efficiency(QE) and collection efficiency (CE)[1], and incident photons with wavelength, area, angle and environment. As known, Quantum efficiency is defined as the process of incident photons that are converted into photoelectrons on the photocathode of PMT through photoelectric process. Then, with the effect of electric field, the photoelectrons are collected by the collector and multiplied by following dynodes where CE is defined to describe the collection fraction. The higher the quantum efficiency and the collection efficiency, the higher detection efficiency and more photoelectrons can be detected. The DE also depends on the PMT structure and the voltage divider [2].

The number of incident photons are difficult to be measured precisely till now and the CE is also hard to measure directly as an internal process. According to the current running and proposed experiments, QE is calibrated by Borexino[3], Daya Bay[4-5] and XENON1T[6] to simulate and guarantee the response of detector, while Double Chooz[2] and IceCube[7] were focusing on the photon detection sensitivity. Double Chooz adopted a relative method to compare the anode signal to a reference tube where CE is 0.9 provided by manufacturer[2]. IceCube applied a calibrated Photodiode to monitor the photon flux and to obtain the relative photon DE of testing tubes[7].

In Daya Bay, 8 inches tubes were used and the QE is measured by the light from a lens and mirror system to cover the whole photocathode[5]. JUNO is a multi-purpose experiment to measure neutrino mass hierarchy with reactor antineutrinos, as well as supernovae, geo-neutrino and atmospheric neutrino and so on [8]. In order to achieve the key physics goals of JUNO, $3\%/\sqrt{E_{vis}}$ energy resolution is the essential requirement[9-10], where 20 thousands 20 inches large-area PMTs, the key sensor in the detector, will be used where DE needs to be better understanding for better detector prediction. While, so far there is no an appropriate method to define and measure the DE of large-area PMTs, which depends on the size and incident angle of light, the parameterization of Mento Carlo simulation. To study the effective response of PMTs and predict the energy response of JUNO central detector, here we are trying to measure CE and QE of PMTs with point light in details to have a better understanding to DE. Hamamatsu Photonics and HZC Photonics have developed a series of dynode PMTs with different size and collection dynode, and it is worth to mention that Hamamatsu 20'' R12860 and HZC 3'' XP72B22 will be used in JUNO.Totally 7 different tubes have been tested to study and compare the relative CE, where the collecting dynodes are very different among the PMTs and more details are described in the Table. 1 and Fig.1.

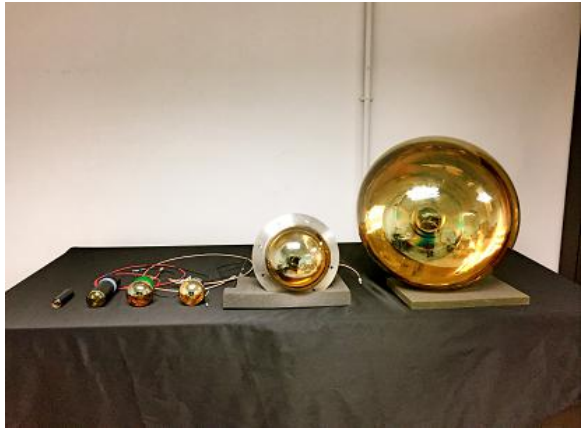

Fig.1: Tested tubes. From left to right:R1355, R7725-100, XP72B22, R12199, R5912 and R12860.

Table. 1: 7 Tubes with different collector electrodes and purpose

| Manufacturer | Serial Number | Size | Collect electrodes |
|---|---|---|---|
| Hamamatsu | R1355 | 1'' | ~1cm*1cm, no wire |
| Hamamatsu | R7725-100 | 2'' | ~1cm*1cm, no wire |
| Hamamatsu | R12199 | 3'' | - |
| Hamamatsu | R5912 | 8'' | ~3cm*3cm, mesh plate |
| Hamamatsu | R12860 | 20'' | ~phi8cm, with mesh |
| HZC | XP72B22 | 3'' | 2cm*2cm,4 wires*4wires |
| HZC | XP72B20 | 3'' | 2cm*2cm,4 wires*4wires |

## 2. System setup

The radiant sensitivity and cathode blue sensitivity are normally used to describe the QE of PMT and they were used by XENON 100 dark matter experiment to calibrate the absolute QE[11]. A relative test to PD with known QE is also a common method to measure QE of PMT[12]. And the pulse height distribution (PHD) of single photoelectron was measured to get the CE of MCP tube[13]. In this paper, point light of 420 nm is adopted and we designed a relative system with a reference tube to obtain the relative QE. At the same time, we set up a system to get average photoelectron considering Poisson distribution to evaluate light intensity proportional to DE and also to avoid non-linearity effect for strong light intensity. The ratio of relative DE and QE is defined as the relative CE of PMT.

PMTs are really sensitive to the wavelength of incident light and most tubes are response to the wavelength from 300 nm to 700 nm. In our test, a given wavelength 420 nm is considered. Besides, the different glass shape requires the incident light spot as small as possible and in vertical to the surface to avoid incident angle effect and electronic field non-uniformity. This configuration is used for the following tests. For further checking, more than one light intensities for QE and relative DE measurements have been implemented to cross check the uncertainty.

## 2.1 Quantum efficiency

The block diagram of QE test system is shown in Fig.2. A light emitting diode (LED) is used to fire the PMT and a Keithley 6485 picoammeter is used to measure the Cathodic Direct Current (DC) from the photocathode to the first dynode which shorted with all the other dynodes.

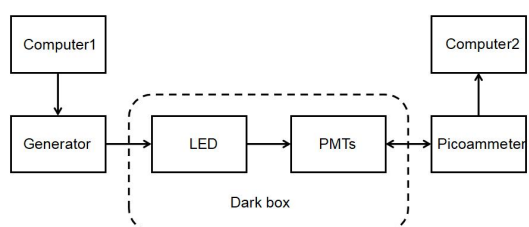
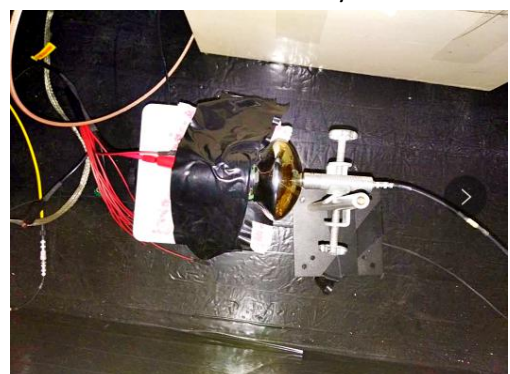

Fig. 2. Experimental setup for QE measurement

The LED and testing PMT are put in a dark box with magnetic shielding. The driver of LED is a generator, which provides the driving voltage to generate pulse light with a fast pulse. The light spot diameter of the LED is set to approximate 3 mm with a focus lens and Vertical incident to the center of the glass, where the LED touches the glass to avoid angle uncertainty. The picoammeter, controlled by another computer, will automatically supply a series voltages to simulate the collection setting, and also measure the current.

For QE, relative measurements to a reference tube is used here. Hamamatsu Photonics has developed a new serial of PMT, R12860, which is stated for high quantum efficiency. R12860-EA0073, the reference tube here, is calibrated with different wavelength by the manufacture and our laboratory, and its QE vs. light wavelength is shown in Fig. 3. Particularly, the QE of tube EA0073 is 25.8% with relative 3% uncertainty at wavelength of 420 nm.

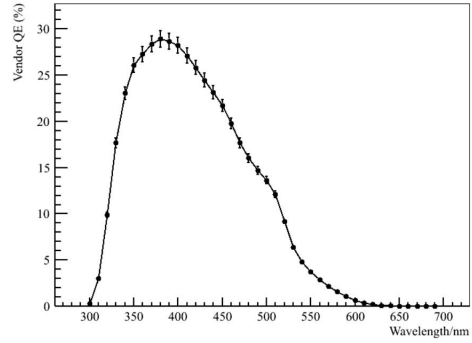

Fig.3. QE of EA0073 vs. light wavelength

As known, QE is proportional to the photocurrent subtracted the dark current when the setting is far from photocathode saturation. Then the QE of the candidate PMT can be obtained by the calculated ratio of the current of the candidate tube and the reference tube by：

$$QE_{testing} = QE_{reference} \times \frac{I_{testing}}{I_{reference}} \qquad (1)$$

Where $QE_{testing}$ is the QE of the testing PMT and $QE_{reference}$ is the QE of reference tube. $I_{testing}$ and $I_{reference}$ are the measured current subtracted the dark current of candidate and reference PMT respectively.

In our design, the 20'' tube R12860-EA0073 is the reference tube which will be tested firstly, then the candidate tube will be tested secondly to avoid instability of the system. The tubes will be tested with the collection voltage from 0 to 800 V in 50 V step. Data acquisition time for each HV point is 5 seconds and collection one measure one second. The averaged value of the five points is defined as the dark current or light current. After subtracting the dark current, we will obtain the current of each HV. Totally 3 light intensities with different LED driving voltage and duty cycle are tested where the duty cycle is the time ratio of the LED driving voltage on to off. The higher driving voltage and duty cycle of LED, the higher light intensity.

These configurations are selected to satisfy the measurement for different tube and dark current noise level. Here driving voltage(2.9 V) -duty cycle(10%), 2.875 V-25% and 2.85 V-50% are measured for the candidate and reference tubes.

The testing of candidate tubes is following the same procedure as the reference tube R12860-EA0073, which has described above. Generally, for a constant incident light intensity, the current increases following the HV increasing because of the collecting efficiency effect. It is obvious that the plateau is stable when HV is higher than 400 V where the current is calculated by the average from 400 V to 800 V. 1.27 nA, 2.10nA and 2.82nA are corresponding to the calculated current of 2.9V -10%, 2.875V-25% and 2.85V-50% of R12860-EA0073 in Fig. 4 where the plateau curve of R12860-EA0073 and XP72B22 are shown. With Formula (1), QE of XP72B22 can be obtained as presented in table.2.

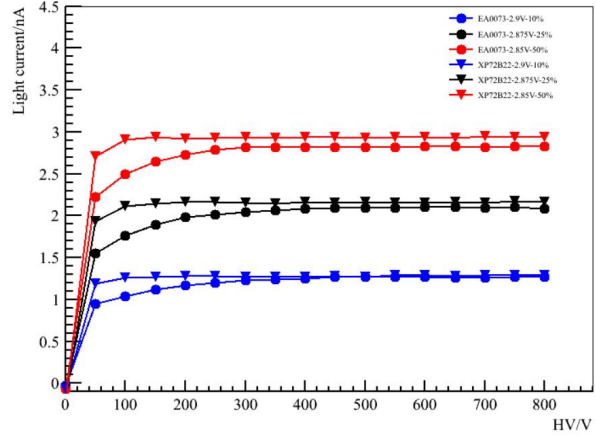

Fig.4. The measured current of EA0073 and XP72B22 vs HV. The red curve is with the configuration (2.85V)-duty ratio (50%), black one is 2.875V-25% and the blue is for 2.9V-10%.

Table.2. The QE of XP72B22

| Driving V-Duty Ratio | 2.900V-10% | 2.875V-25% | 2.850V-50% |
|---|---|---|---|
| EA0073-I/nA | 1.27 | 2.10 | 2.82 |
| XP72B22-I/nA | 1.28 | 2.16 | 2.93 |
| Ratio | 1.008 | 1.029 | 1.039 |
| Averaged-Ratio/Sigma | | 1.025/0.016 | |
| QE/Sigma | | 26.45%/1.21% | |

## 2.2 Relative photon detection efficiency

The relative photon detection efficiency testing system, shown in Fig.5, is based on the charge measurement with QDC measurement (CAEN V965). The generator is controlled by computer to generate a pulse with 1 kHz frequency and 20 ns width to drive the LED as light source, where the light is split into two by fiber to fire the monitoring and testing PMTs. The diameter of the final light spot with 420 nm wavelength is about 1 mm and incident to the center of PMTs' photocathode vertically. In order to monitor the light stability, a 3" tube XP72B20 is used as monitoring tube. All the tubes are tested with magnetic shielding. The output anode signal is input to the QDC with 300 ns gate synchronized with the driving pulse, converted into charge and record by a computer[14]. The charge constant of the QDC is 25 fC/channel.

Considering the whole process of photoelectric effect on photocathode and photoelectrons' collection by the first dynode, the signal distribution from the anode follows the Poisson distribution illustrated as:

$$P(n,\mu) = \frac{\mu^n e^{-\mu}}{n!} \quad (2)$$

Where $\mu$ is the mean photoelectrons collected by the dynode, which can be calculated by the signal count, and $P(n,\mu)$ is the probability that n photoelectrons will be observed when their mean is $\mu$. In other words, if we have a stable light as input,, the ratio of mean photoelectrons from the candidate and reference tubes is proportional to the DE ratio.

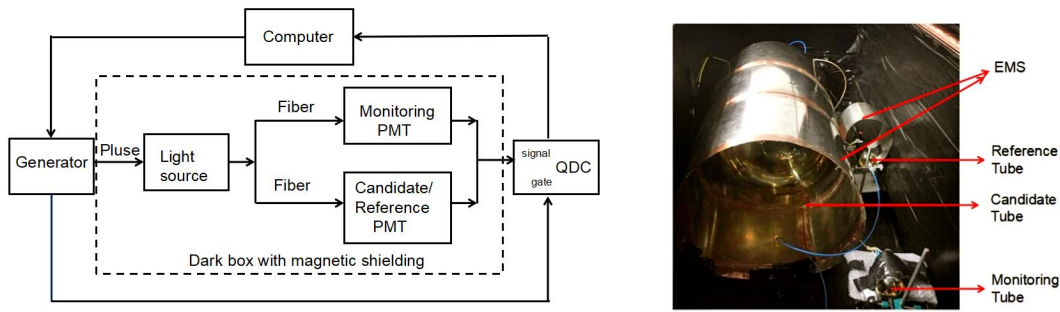

Fig. 5. The scheme of relative photon detection efficiency system (left); The right one is the dark box for relative DE testing system including monitoring tube, and candidate tube with magnetic shielding.

The tubes need to be cooled down and working at 1E7 gain, where the dark count rate needs to be stable and will be subtracted and considered for the uncertainty during the measurements to avoid coincidence bias. A time-dependence dark count rate of XP72B22 is shown in Fig.6. After 90 minutes, the dark count is stable to meet the requirement of variation <1% (relative) bias. A threshold at 0.25 p.e are selected to obtain the count of signal. The reference tube for DE measurement need to be tested before and after each candidate tube to check the system stability. We select a smaller tube (R12199) as the reference tube instead of 20'' R12860-EA0073, it is more convenient to save time and make the testing system more stable.

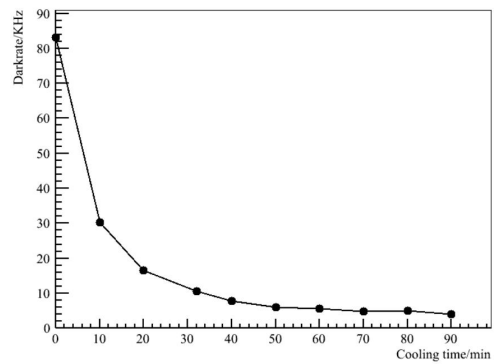

Fig.6 The time-dependence of dark count of XP72B22

Taking the reference tube R12199 as an example, we set its operating HV at 1230V for 1E7 gain. The mean of pedestal in charge spectrum is 1325 channel, and the threshold of 0.25 p.e. is 1341 channel. To double check the system and validate the effect of dark count, several light intensities were tested listed in LED driving configuration 0V-1K-0.002% (driving voltage-frequency-duty cycle ), 4.5V-1k-0.002% and 4.7V-1k-0.002% and the measured charge spectrum are depicted in the lift of Fig.7, where the driving voltage and duty cycle have been described in section2.2. The statistics of pedestal, signal and mean photoelectrons are listed in the Table.3. The mean photoelectron is the calculated number subtracted with dark count. XP72B20 is used to monitor the stability of light intensity to correct the variation of light intensity. The data of XP72B22 is shown as an example comparing to R12199 including the monitoring tube XP72B20 in table. 3. From the table if we assume that the DE of R12199 is 100% at 420 nm wavelength, the relative DE of XP72B22 measured with the LED driving voltage 4.5V and 4.7V can be obtained from formula (3) and formula (4) separately, therefore the measured relative DE of XP72B22 to R12199 is 101.72% from the average of these light intensities.

$$\frac{0.3872/0.2033}{0.3853/0.2047}*100\% = 101.18\% \qquad (3)$$

$$\frac{1.2971/0.6336}{1.2884/0.6435}*100\% = 102.25\% \qquad (4)$$

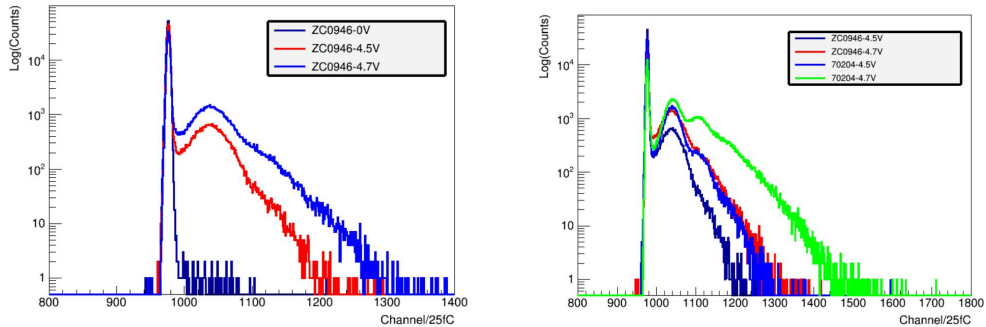

Fig.7. left: the charge spectrum of reference tube R12199 under dark, 4.5V and 4.7V of LED drving voltage. Right: the charge spectrum comparison of reference tube R12199 and the testing tube XP72B22 in 4.5V and 4.7V of LED driving voltage.

Table. 3. The measurements of Candidate tube XP72B22, Reference tube R12199 and the monitoring tube XP72B20.( Total counts is the total number of events while the Signal Counts means the pulses over the threshold. Mu is the mean photoelectrons and Sub_mu is the results with light subtract the mu in the dark)

| Candidate/ Reference | Total Counts | Signal Counts | Mu | Sub_mu | Monitoring Tube | Total Counts | Signal Counts | Mu | Sub_mu |
|---|---|---|---|---|---|---|---|---|---|
| XP72B22 Dark | 284350 | 345 | 0.0012 | | XP72B20 Dark | 284009 | 189 | 0.0003 | |
| 4.5V | 285233 | 91803 | 0.3884 | 0.3872 | 4.5V | 286720 | 63628 | 0.2036 | 0.2033 |
| 4.7V | 284671 | 206955 | 1.2983 | 1.2971 | 4.7V | 283319 | 150327 | 0.6339 | 0.6336 |
| R12199 dark | 284922 | 202 | 0.0007 | | XP72B20 dark | 285428 | 169 | 0.0003 | |
| 4.5V | 283532 | 90800 | 0.3860 | 0.3853 | 4.5V | 282876 | 60003 | 0.2050 | 0.2047 |
| 4.7V | 284052 | 205788 | 1.2891 | 1.2884 | 4.7V | 284088 | 148191 | 0.6438 | 0.6435 |

## 3 Relative CE of different tubes

All the tubes are measured with LED configurations: 2.900V-10%, 2.875V-25%, 2.800V-50% and without light (dark current) to get QE. The measured QE is calculated from the averaged DC comparing to the reference tube, which also checked with the vendor QE as listed in Table.4, showing consistent results. All the measured QE of all the tested tubes are listed in Table.5.

Table.4. QE comparison with Vendor data

| PMT_ID | Ratio | QE(%) @420nm Average/Sigma | Vendor QE (%)@420nm/ Value/Sigma |
|---|---|---|---|
| R12860-EA0073 （reference） | 1 | 25.8/0.77 | 25.8/0.77 |
| R1355(testing) | 0.86 | 22.29/0.69 | 21.80/- |
| R7725-100(testing) | 1.48 | 38.20/1.82 | 38.17/- |

Normally it is known that DE is mainly contributed by QE and CE, and here we are assuming DE=QE×CE following our configuration. With the measured relative QE and DE of different tubes, relative CE could be obtained referring to R12199, shown in Table.5. totally six R12860 Hamamatsu tubes are measured to compare the average relative CE with other types of tubes, listed in Table.6.

Table.5 CE of different tubes normalized to R12199

| Manufacturer | Serial Number | QE(%)/ Sigma(%) | DE (%)/ Sigma(%) | CE (%)/ Sigma(%) |
|---|---|---|---|---|
| Hamamatsu | R12199 | 25.52/1.01 | 100/1.00 | **100/4.08** |
| Hamamatsu | R1355 | 22.29/0.69 | 86.67.00/2.15 | **99.21/3.73** |
| Hamamatsu | R7725-100 | 38.20/1.82 | 137.38/0.19 | **91.77/4.37** |
| Hamamatsu | R5912 | 24.27/0.78 | 89.41/0.68 | **94.01/3.08** |
| Hamamatsu | R12860 | 25.80/0.77 | 108.89/0.89 | **107.69/3.37** |
| HZC | XP72B22 | 26.45/0.89 | 101.72/0.76 | **98.10/3.39** |

Table.6 CE of Hamamatsu R12860 referring to R12199

| Manufacturer | Serial Number | PMT_ID | CE (%)/ Sigma(%) |
|---|---|---|---|
| Hamamatsu | R12860 | EA0302 | 96.26/3.01 |
| Hamamatsu | R12860 | EA0473 | 103.24/3.23 |
| Hamamatsu | R12860 | EA0367 | 102.74/3.21 |
| Hamamatsu | R12860 | EA0743 | 106.91/3.34 |
| Hamamatsu | R12860 | EA0073 | 107.69/3.37 |
| Hamamatsu | R12860 | EA0111 | 106.60/3.34 |
| | Average/RMS | | **103.91/3.25** |

## 4.Conclusion and discussion

In this paper, we set up a relative QE and DE measurement system to obtain the relative QE and DE for the comparison on the relative CE of six different dynode tubes under the specified configuration, where multiple light intensities are used to cross check the uncertainty. It is consistent in relative QE and DE measurement with multiple light intensities, and at the same time the errors of the corresponding parameters of each tube were given. Based on the measurements Hamamatsu R12860 shows the maximum collection efficiency, and Hamamatsu R12199, R1355 and HZC XP72B22 also shows similar level within the uncertainty. The CE difference of tubes might be source from the different structure of the internal collection electrodes of the tubes. If we assume that the measured relative CE as the absolute reference point, then we will get the definition for absolute DE with QE measurements. At the same time, following the configuration about incident light and DE definition, we can further define the DE of a large area PMT with similar algorithm as discussed in the reference[15], which will be used for detector simulation in JUNO.